\documentclass[fleqn,usenatbib]{mnras}

\usepackage{mathptmx}

\usepackage[T1]{fontenc}

\DeclareRobustCommand{\VAN}[3]{#2}
\let\VANthebibliography\thebibliography
\def\thebibliography{\DeclareRobustCommand{\VAN}[3]{##3}\VANthebibliography}

\usepackage{graphicx}	
\usepackage{amssymb}	
\usepackage{amsmath}	
\usepackage{pbox}

\title[PSR J2021$+$4026 search for radio emission]{A renewed search for radio emission from the variable $\gamma$-ray pulsar PSR J2021$+$4026}

\author[B. Shaw et al.]{B. Shaw,$^{1}$\thanks{E-mail: benjamin.shaw@manchester.ac.uk}
B. W. Stappers,$^{1}$
P. Weltevrede,$^{1}$
C.A. Jordan$^{1}$
M.B. Mickaliger$^{1}$
\newauthor and A.G. Lyne$^{1}$
\\
$^{1}$Jodrell Bank Centre for Astrophysics, School of Physics and Astronomy, University of Manchester, Manchester, UK, M13 9PL\\
}

\date{Accepted XXX. Received YYY; in original form ZZZ}

\pubyear{2023}

\begin{document}
\label{firstpage}
\pagerange{\pageref{firstpage}--\pageref{lastpage}}
\maketitle

\begin{abstract}
We undertake the first targeted search at 1.5 GHz for radio emission from the variable $\gamma$-ray pulsar PSR J2021$+$4026.  This radio-quiet pulsar assumes one of two stable $\gamma$-ray emission states, between which it transitions on a timescale of years. These transitions, in both $\gamma$-ray flux and pulse profile shape, are accompanied by contemporaneous changes to the pulsar's spin-down rate.  A number of radio pulsars are known to exhibit similar correlated variability, which in some cases involves an emission state in which the radio emission ceases to be detectable. In this paper, we perform a search for radio emission from PSR J2021$+$4026, using archival radio observations recorded when the pulsar was in each of its emission/spin-down states.  Using improved techniques, we search for periodic radio emission as well as single pulse phenomena such as giant radio pulses and RRAT-like emission. Our search reveals no evidence of radio emission from PSR J2021$+$4026. We estimate that the flux density for periodic emission from PSR J2021$+$4026 does not exceed 0.2 mJy at this frequency. We also estimate single-pulse flux limits for RRAT-like bursts and giant radio pulses to be 0.3 and 100 Jy respectively. We discuss the transitioning behaviour of PSR J2021$+$4026 in the context of pulsar glitches, intermittent pulsars and the increasingly common emission-rotation correlation observed in radio pulsars.  

\end{abstract}




\section{Introduction}

PSR J2021$+$4026 is a young isolated neutron star located within the supernova remnant G78.2+2.1 \citep{aaa+09a}. The pulsar has a rotation period of 265.3 ms and a period derivative of $5 \times 10^{-14}$ \nolinebreak s \nolinebreak s\textsuperscript{-1}\citep{rkp+11}.  These spin parameters imply a comparatively high spin-down energy loss rate  $\dot{E} \sim 10^{35}$ ergs s\textsuperscript{-1}, and a characteristic spin-down age $\tau_{\mathrm{char}} \sim 77$ kYr. Position, timing and derived properties of the pulsar are shown in Table \ref{Table:pulsarpars}. 

$\gamma$-ray emission from PSR J2021$+$4026 was first identified using the EGRET\footnote{The Energetic Gamma Ray Experiment Telescope} GeV instrument on-board the Compton $\gamma$-ray Observatory (CGRO) by \cite{lm97} who designated it GEV J2020$+$4023/3EG J2020+4017. It was also detected as a source of pulsed $\gamma$-ray emission during a blind frequency search by the Large-Area Telescope (LAT) on-board the \emph{Fermi} $\gamma$-ray observatory \citep{aaa+09a}.  An X-ray counterpart to PSR J2021$+$4026 was identified  by \cite{thc10} using data from the \emph{XMM-Newton} serendipitous source catalogue. It was designated 2XMM J202131.0$+$402645, and subsequently shown to be a source of pulsed X-ray emission with a period consistent with that of J2021$+$4026 \citep{lhh+13}. Despite previous searches (\citealt{bwa+04}, \citealt{rkp+11}), no pulsed radio emission has been observed from PSR J2021$+$4026. The pulsar belongs to a small class of sources generally regarded as \emph{Geminga-like} (e.g., \citealt{lin16}), named after Geminga (PSR B0633$+$17 / J0633$+$1746) which was the first radio-quiet pulsar known to show pulsed emission at high energies and, like PSR J2021$+$4026, a substantial un-pulsed component.  Observations of such pulsars are invaluable for understanding the physics that underpins pulsar emission and providing constraints on emission geometry. 

\begin{center}
\begin{table}

\caption{Table of measured and inferred parameters of PSR J2021$+$4026. Quantities marked with a $\dagger$ are taken from the ATNF pulsar catalogue \citep{mhth05}.}
\begin{tabular}{ | l | r |}
\hline
\hline
Pulsar parameters &  \\ \hline
Right Ascension, $\alpha$ (J2000) & 20:21:30.733 \\
Declination, $\delta$ (J2000) & +40:26:46.040 \\
Spin frequency, $\nu$ (Hz) & 3.7689908200(7) \\
Spin frequency derivative, $\dot{\nu}$ (Hz s\textsuperscript{-1}) &  $-8.166(2) \times 10^{-13}$ \\
Period epoch (MJD) & 56063.0 \\
Characteristic age, $\tau_{\mathrm{char}}$ (kYr) & 77\textsuperscript{$\dagger$} \\
Spin-down luminosity, $\dot{E}$ (ergs s\textsuperscript{-1}) & $10^{35\dagger}$ \\
Distance, $d$ (kpc) & 2.150\textsuperscript{$\dagger$} \\
\hline
\end{tabular}
\label{Table:pulsarpars}
\end{table}
\end{center}

PSR J2021$+$4026 is known to exhibit correlated emission and spin-down variability (\citealt{abb+13}, \citealt{znl+17}, \citealt{twl+20}). This is a phenomenon that is observed in an increasing number of radio pulsars in which the radio emission state undergoes a transition contemporaneously with changes to the spin-down rate of the star. The emission changes may be characterised by changes to the shape of the pulse profile (so called mode-switching\footnote{often also referred to as mode-changing or simply, moding}), changes to the radio flux density or, like in the cases of the intermittent pulsars, a complete cessation of detectable radio emission (see e.g., \citealt{klo+06}, \citealt{lhk+10}, \citealt{bkj+15}, \citealt{lsf+17}, \citealt{ssw+22}). In many of these cases, radio pulsars are observed to assume two metastable states which they transition between on a quasi-regular basis. There are comparatively few examples of such state-changing behaviour in radio-quiet, high-energy emitting pulsars. The spin-down rate of the radio-quiet X-ray / $\gamma$-ray pulsar B0540$-$69 underwent a single, large (36\%) increase in 2015 \citep{mgh+15} but no radiative effects associated with this change have been observed to date. More recently \cite{gyl+20} reported repeated bi-modal spin-down transitions in the $\gamma$-ray pulsar PSR J1124$-$5916.  PSR J2021$+$4026 is the only $\gamma$-ray pulsar known whose spin-down transitions are accompanied by detectable changes to its $\gamma$-ray profile.

\begin{figure*}
    \centering
    \includegraphics[width=1.9\columnwidth]{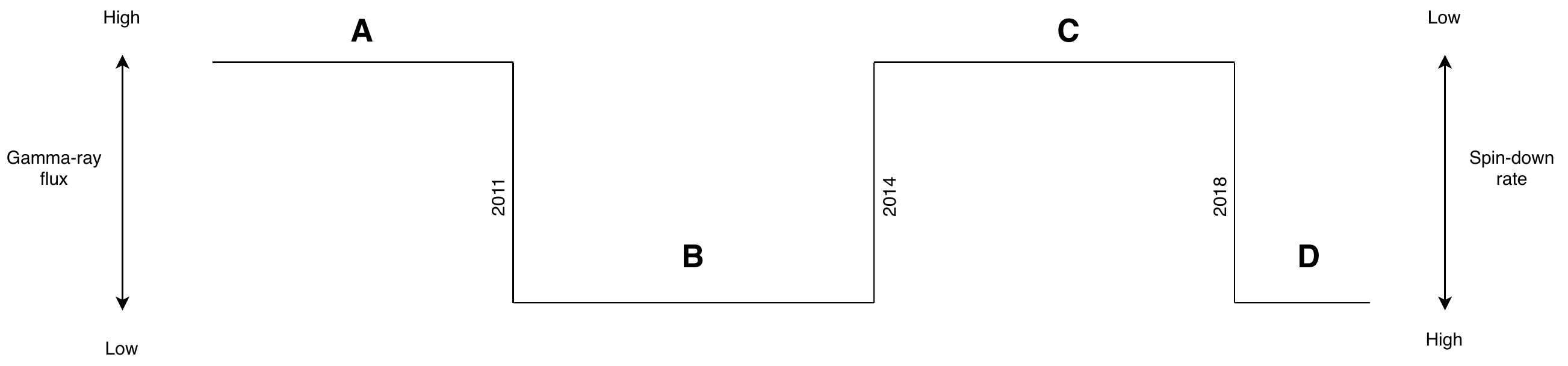}
    \caption{Schematic representation of the transitions in $\gamma$-ray flux and spin-down rate of PSR J2021$+$4026, covered by our dataset, between 2011 and 2020 inclusive. A later transition to a high $\gamma$-ray flux state has since occurred (see text). A downward deflection represents an increase in the magnitude of the spin-down rate. Note that despite the depiction of the transitions as equal in magnitude, there is some variation in the jump sizes in both spin-down rate and $\gamma$-ray flux (see text).}
    \label{fig:state_model}
\end{figure*}

Figure \ref{fig:state_model} shows schematically the transition timeline of PSR J2021$+$4026 from 2011 to 2020.  The first transition was observed by \cite{abb+13} wherein an 18 per cent drop in $\gamma$-ray flux occurred in 2011 October (MJD 55850) as well as the disappearance of faint bridge emission between the two primary peaks of the pulsar's double-peaked $\gamma$-ray profile (see also \citealt{ntc16} and \citealt{znl+17}). This occurred coincidentally with an $\sim$6 per cent increase in the magnitude of the spin-down rate.  In 2014, approximately 38 months later, the pulsar transitioned back towards its pre-2011 state when the $\gamma$-ray flux increased by $\sim$11 per cent and the profile returned to its pre-2011 shape, along with an $\sim$6 per cent decrease in the spin-down rate \citep{znl+17}. In February 2018, PSR J2021$+$4026 transitioned once again with an $\sim$13 per cent drop in $\gamma$-ray flux and an $\sim$3 per cent increase in the spin-down rate \citep{twl+20}. \cite{pm23} showed that the pulsar returned to the high $\gamma$-ray flux state in June 2020.  \cite{wth+18} studied the evolution of the X-ray emission properties of PSR J2021$+$4026 using data acquired by \emph{XMM-Newton}, finding no significant differences between the states. This was corroborated by \cite{rmt+21} who further inferred that the X-ray emission originates from a single extended hotspot.

At Jodrell Bank Observatory (JBO) we have a number of archival observations of PSR J2021$+$4026, undertaken whilst the pulsar was in each of its spin-down/emission states. Previous unpublished analyses of these data have not revealed any evidence of pulsed radio emission.  We revisit these observations and utilise improved techniques to carry out a renewed search for periodic pulsations from PSR J2021$+$4026 at 1.5 GHz. 

In addition to the periodic radio emission that is characteristic of the `normal' pulsar population, an increasing number of neutron stars are seen to emit more sporadic single pulse behaviour. Rotating RAdio Transients (known as RRATs) \citep{mll+06} emit only a small number of detectable individual pulses per hour and so are typically not discovered using periodicity search techniques.  With a sufficient number of detected pulses, the application of standard pulsar timing methods reveals underlying periodicities and spin-down rates which are consistent with those of the normal pulsar population (see \citealt{kkl+11} for a review), supporting the association with neutron stars. Although the mechanism behind RRAT emission is not well understood, they may represent extreme examples of the intermittent pulsars (e.g., \citealt{klo+06}, \citealt{wmj07}, \citealt{lsf+17}), which spend extended periods of time in which no radio emission is detectable.   

A small, but growing, number of pulsars are known to emit giant radio pulses\footnote{Often referred to simply as \emph{giant pulses}} (GPs). These events are characterised by sporadic bursts of radio emission whose flux densities are in significant excess of the single-pulse average (see e.g., \citealt{jr04a}, \citealt{msb+19}). They also typically occupy a smaller proportion of a rotation than a pulsar's normal periodic radio pulsations.  To date, only a handful of pulsars are seen to emit GPs and these sources tend to exhibit strong high-energy emission and have extremely strong magnetic flux densities at the light cylinder. For example, PSR B0540$-$69 is a prolific emitter of GPs \citep{fak15} and although the magnetic flux density at the light cylinder of PSR J2021$+$4026 is modest, a detection of GPs would provide unique insights into pulsar emission physics. For these reasons we also undertake a separate search for any transient radio emission from PSR J2021$+$4026.  

The paper is set out as follows. We provide a description of our observations and data recording strategy in \S \ref{observations} and our data analysis techniques for the periodic and single-pulse searching in \S \ref{analysis}. \S \ref{results} outlines the results of our searches and we discuss their implications in \S \ref{discussion}.

\section{Observations}
\label{observations}

Our dataset comprises sets of observations of PSR J2021$+$4026, from 2011 through 2020, all of which were carried out using the 76-m Lovell telescope at the Jodrell Bank Observatory. In Table \ref{Table:obslist} we list the dates, lengths and frequency parameters of the observations as well the specific $\gamma$-ray flux/spin-down state the pulsar occupied in each of the observations as reported in \cite{twl+20} and shown in Figure \ref{fig:state_model}. Observations were made of PSR J2021$+$4026 when it was in high $\gamma$-ray flux/low spin-down (HGF/LSD) state A and low $\gamma$-ray flux/high spin-down (LGF/HSD) states B and D. No observations were made of the pulsar during the HGF/LDS state C. Our data exists in the form of \emph{time series} (TS) filterbanks (time-frequency planes) for observations 4$-$8 and \emph{pulsar fold} (PF) folded archives for observations 1$-$3. 


\begin{center}
\begin{table}
    \caption{Dates, data formats (see text), durations $\tau_{\mathrm{obs}}$, central observing frequencies $f_c$ and bandwidths $\Delta f$ for all observations of PSR J2021$+$4026 carried out with the Lovell telescope. The rightmost column shows the emission state at the time of each observation, labelled according to Figure \ref{fig:state_model}.}
    \begin{tabular}{ | l | l | l | r | r | r | r | r }
    \hline
    \pbox{0.2cm}{\#} & \pbox{1cm}{Date} & \pbox{0.7cm}{Data \\ Format} & MJD &  \pbox{0.5cm}{$\tau_{\mathrm{obs}}$ \\ (mins)} & $f_c$ & $\Delta f$ & State \\ \hline
    1 & 2011-07-20 & PF & 55762 & 433 & 1532 & 384 & A \\
    2 & 2011-07-22 & PF & 55764 & 453 & 1532 & 384 & A \\
    3 & 2011-07-26 & PF & 55768 & 370 & 1532 & 384 & A \\
    \hline
    4 & 2013-11-19 & TS & 56615 & 124 & 1532 & 400 & B \\
    5 & 2013-11-20 & TS & 56616 & 514 & 1532 & 400 & B \\
    6 & 2013-11-29 & TS & 56625 & 62 & 1532 & 400 & B \\
    7 & 2014-01-12 & TS & 56669 & 60 & 1532 & 400 & B \\
    \hline
     8 & 2020-02-21 & TS & 58900 & 61 & 1532 & 336 & D
    \end{tabular}
    \label{Table:obslist}
\end{table}
\end{center}

Filterbanks for each of the TS observations are formed as follows. Using a polyphase filter, the 400 MHz wide band is channelised into 25 channels, each of which has a width of 16 MHz and a sampling time of (1/16 MHz) s. Using {\sc digifil} from the {\sc dspsr} software suite \citep{vb11}, each of these channels is then more finely channelised into 32 channels resulting in a total of 800 channels, each with a width of 500 kHz. Each sub-band is resampled to a sampling time of of 256 $\mu$s. Each filterbank has a maximum duration $\tau_{\mathrm{obs}} \approx 31$ minutes.  

In some observations, the channels at the edges of  the bandpass are removed due to bright, terrestrial radio frequency interference (RFI) at those frequencies resulting in a useable bandwidth that is slightly less than 400 MHz. The centre frequencies and final bandwidths are listed in Table \ref{Table:obslist}. 

\begin{figure*}
    \centering
    \includegraphics[width=\columnwidth]{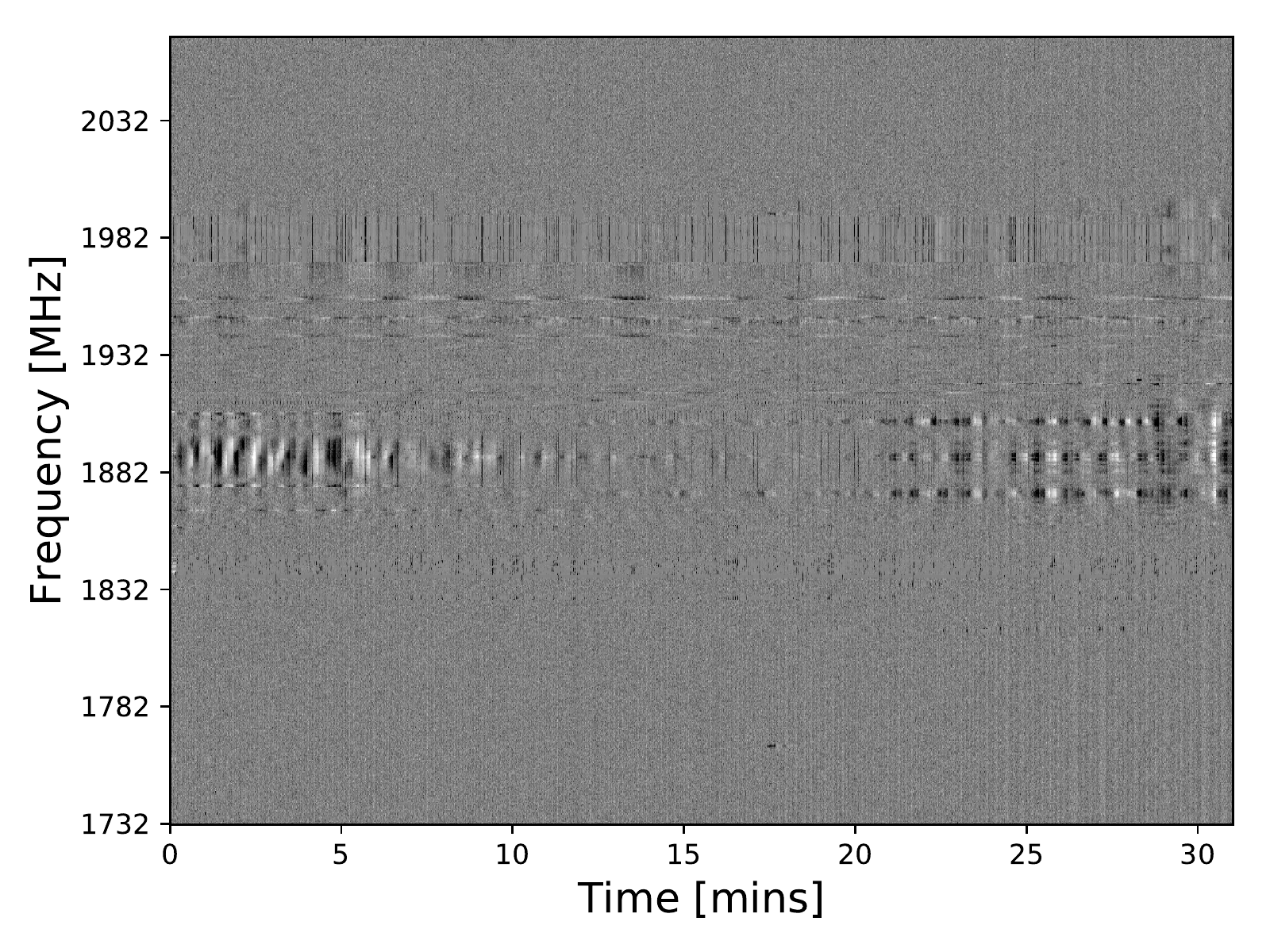}
    \includegraphics[width=\columnwidth]{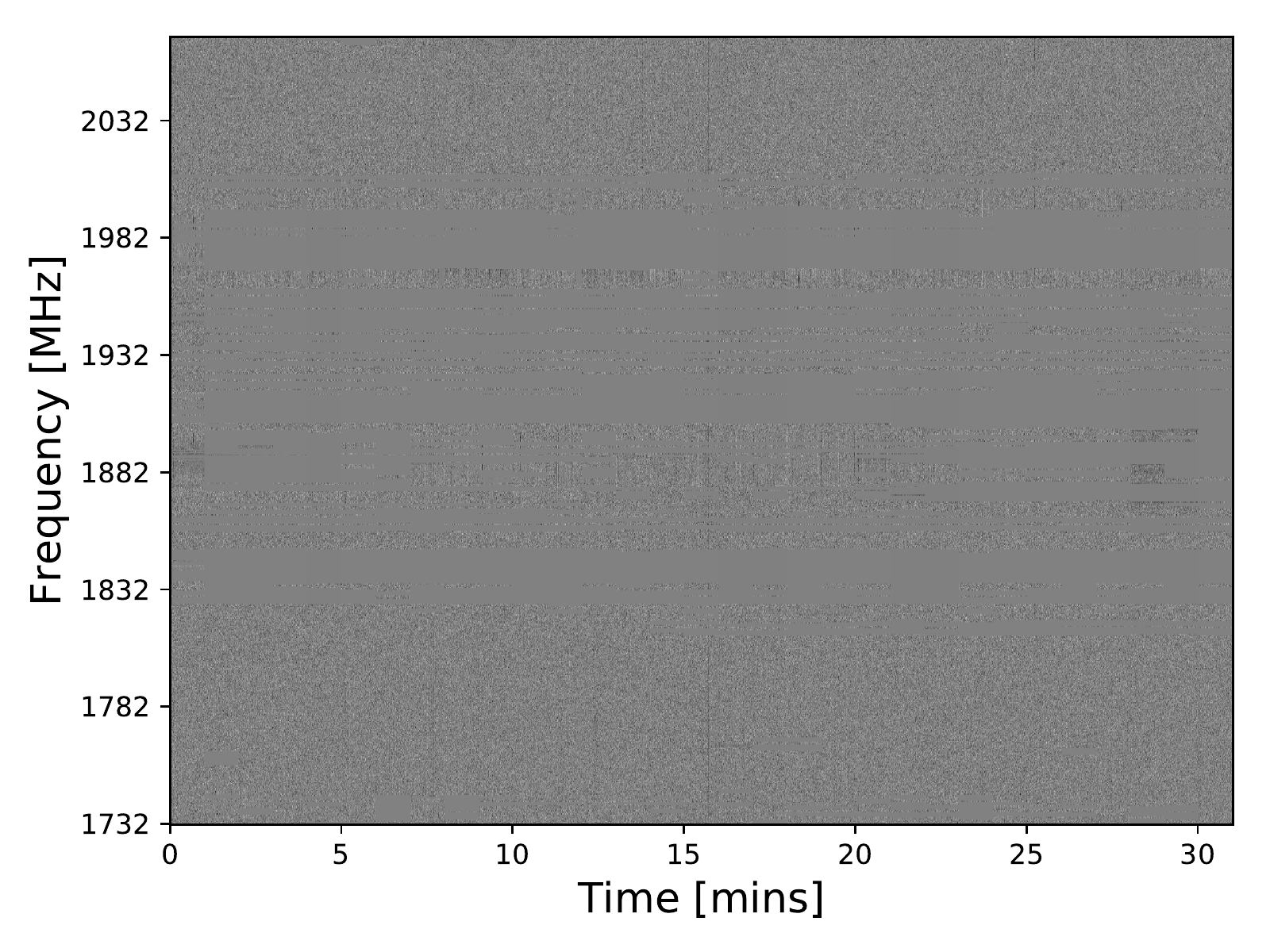}
    \caption{Left: The raw dynamic spectrum of the first 31 minutes of observation number 8 (see Table \ref{Table:obslist}). Right: The same data again after our RFI mitigation techniques have been applied (see text).}
    \label{fig:rfi-spectra}
\end{figure*}

In order to remove the effects of strong periodic RFI from our TS data (see Table \ref{Table:obslist}), samples from each 500 kHz channel were read as individual time series. For each of these time series we compute the fast-Fourier transform (FFT)\footnote{using the {\sc fft} method in the {\sc Numpy} suite. See https://numpy.org/.}. Fourier bins from the resulting power spectrum, whose power values exceed a threshold of 50,000 are adjusted to the median value of the whole power spectrum. The power spectra are individually inspected to ensure that peaks that occur around the known rotation frequency of the pulsar, and any associated harmonics, are not readjusted. The adjusted power-spectrum is then transformed back to a time series by means of in inverse-FFT. Discrete, narrowband instances of RFI are excised using the inter-quartile range RFI mitigation algorithm {\sc iqrm\_apollo}\footnote{https://gitlab.com/kmrajwade/iqrm\_apollo} \citep{mor21}.  Figure \ref{fig:rfi-spectra} shows the effect of applying both of these techniques to a 31-minute chunk of observation number 8.

The filterbanks for our PF observations were formed using the same channelisation method as described above. A median filtering algorithm was employed to remove RFI.  Due to the presence of free electrons in the interstellar medium, radio pulsar signals undergo dispersion as they propagate from the source to the observer such that signals at lower frequencies are delayed with respect to those at higher frequencies. The process of dedispersion corrects this by applying a frequency dependent offset to each channel according to the known \emph{dispersion measure} (DM) of the pulsar (see next section for a discussion of the DM along the line-of-sight to PSR J2021$+$4026). The data from observations 1$-$3 were dedispersed, using a single trial DM value, and folded at the pulse period onto 256 bins and stored to disk as PF data archives. The originally filterbanks have since been removed from our database prior to the commencement of this work, requiring us to undertake a distinct search method using the PF data. These methods are described in the next section.  

\section{Analysis}
\label{analysis}

As PSR J2021$+$4026 has not previously been detected at radio frequencies, its DM is not known a-priori. However, an estimate of the DM can be obtained using a model of the Galactic free electron density at the position and distance of the pulsar. Using HI absorption spectra, \cite{lgr13} determined an association between G78.2+2.1 and the $\gamma$-Cygni nebula, for which \cite{bdw78} computed the range of possible distances to be 1.7$-$2.6 kpc.  The Cordes-Lazio NE2001 model\footnote{see https://www.nrl.navy.mil/rsd/RORF/ne2001/} \citep{cl2003}, taking the distance to PSR J2021$+$4026 to be 2.15 kpc (the centrepoint of the distance range to G78.2+2.1), predicts a DM of $\sim$25 pc cm\textsuperscript{-3}. However, this value ignores the possible contribution to DM from the region of dense nebulosity in which the pulsar resides (see \citealt{jsb+09} for a discussion of other pulsars in the Cygnus region) therefore it is necessary to search a much wider range of DM values than that predicted from models of the interstellar medium alone. Using the {\sc sigproc} tool {\sc dedisperse}, each 31 minute-long TS filterbank is dedispersed at DM trial values ranging from $0 \leq \mathrm{DM} \leq 500$ with a step size $\Delta \mathrm{DM} = 2$, forming a total of 251 dedispersed time series for each observation.    

\subsection{Pulsar Search (PS) data}

In order to search for periodic signals from PSR J2021$+$4026 using our TS datasets,  we use {\sc riptide}\footnote{https://github.com/v-morello/riptide} \citep{mbs+20} which combines a fast-folding algorithm with a matched filtering technique. We process each of the 31 minute long chunks of TS data individually. The {\sc riptide} tool {\sc rffa} takes a set of dedispersed time series and folds each at a large number of closely spaced trial periods. When the time series is folded at the period corresponding to the pulsar's rotation period, a pulse will arise if radio emission is detectable. For each trial period, the folded time series is convolved with a set of boxcar functions with varying width in order to optimally extract the signal-to-noise ratio (S/N) of any detected pulsed emission. In this case the boxcars widen by a factor of 1.5 for each convolution. Peaks with a S/N greater than 7 are designated as candidates. The resulting data products include the parameters of all detected candidate signals (e.g., period, DM, S/N, pulse width, etc) as well as diagnostic plots, such as periodograms, which can be manually inspected to determine which, if any, are likely to be of genuine astrophysical origin. 

A precise rotation period of 265.3 ms for PSR J2021$+$4026 has been determined for PSR J2021$+$4026 from high energy observations (e.g., \citealt{rkp+11}) which narrows down the search parameter space. The lowest period expected is calculated by extrapolating the rotational parameters in Table \ref{Table:pulsarpars} one year back to the earliest epoch at which we have an observation (observation 1 in Table \ref{Table:obslist}). Due to the variable nature of the rotation of the pulsar we are not able to use this ephemeris to accurately extrapolate forwards in time to the epoch of most recent observation (observation 8). Instead we extrapolate forwards from the ephemeris from \cite{twl+20} (see Table 1, therein) which describes the rotation of PSR J2021$+$4026 following the most recent transition.  We search over a small range of trials periods encompassing the known period, ranging from 262.5 ms to 267.5 ms. For the purposes of candidate viewing, the time series are folded onto 256 phase bins divided into 32 subintegrations and stored to disk.

In order to search for single pulse events from PSR J2021$+$4026 in our TS data we utilise the {\sc Heimdall} software package\footnote{https://sourceforge.net/projects/heimdall-astro/}. {\sc Heimdall} searches for single-pulse events in time series generated from a range of trial DM values. Here we used trial DM values from 0 to 500 pc cm\textsuperscript{-3}.  Single pulse candidates produced from {\sc Heimdall} are then classified, as likely to be astrophysical or otherwise, by the {\sc Fetch} machine learning package \citep{aab+20}. Compelling single-pulse candidates are then verified by manual inspection. 

\subsection{Pulsar Fold (PF) data}

Each individual PF archive file contains raw data folded onto 1024 phase bins at the pulse period and dedispersed with a trial DM of 100 pc cm\textsuperscript{-3}. Time and frequency resolution is partially preserved in this process with each archive containing one pulse profile for each of 50 frequency bands and up to 31 (1-minute long) subintegrations.  

We use the program {\sc pdmp} from the {\sc psrchive} package \citep{hvm04} to search each archive for an optimal period and DM around values of the initial period and DM. We search a period range of 20 $\mu$s centred on the original folded value of 265.338 ms. The initial DM trial value of 100 pc cm\textsuperscript{-3}  was chosen in order to minimise the residual smearing of the pulse, due to the finite frequency channel bandwidths, if dedispersed at the wrong value which, if greater than the pulse width, will have a detrimental effect on the measured S/N.  Using the period$-$width relation of \cite{jk19}, we estimate the full width at 10 per cent maximum of any radio pulses from PSR J2021$+$4026 to be $\sim$10 ms, assuming a single-component profile.  We search a DM range of 200 pc cm\textsuperscript{-3} centred on starting DMs of 100 pc cm\textsuperscript{-3} and 250 pc cm\textsuperscript{-3}.  In other words, each archive is searched twice, once for each starting DM. The resultant maximum smearing of any radio pulsations due to an incorrect trial DM is 3 ms and 7 ms respectively. Each optimisation returns the values of the period and DM that produce the highest signal-to-noise ratio and a set of diagnostic plots which are manually inspected for evidence of radio pulsations.


%
%
%


\section{Results}
\label{results}

\begin{figure*}
    \centering
    \includegraphics[width=2.1\columnwidth]{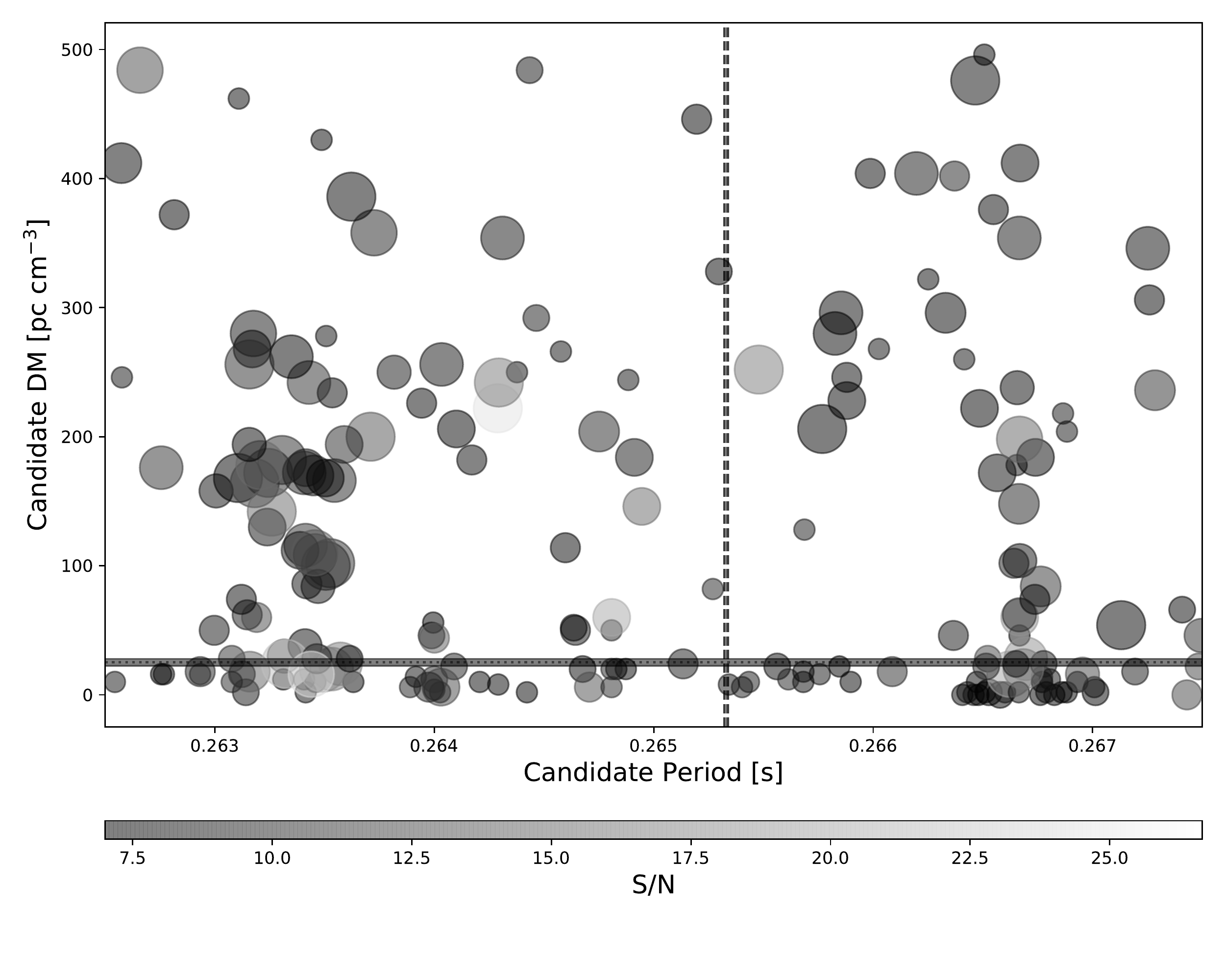}
    \caption{Period$-$DM distribution of all periodicity candidate detections of PSR J2021$+$4026 from observations 4 to 8 generated with {\sc riptide}. Each data point is assigned a shade according to the signal-to-noise ratio of the candidate and a size according to the relative duty cycle of the detected pulse with values ranging from 0.4\% to 16.4\% of one pulse period. The horizontal dashed line represents the predicted value of the pulsar's dispersion measure (see text). The shaded area represents the $3\sigma$ uncertainty of the predicted DM. The vertical dashed lines represents the known values of the period of PSR J2021$+$4026 at the start and end of our dataset. The $3\sigma$ uncertainty regions of the period are plotted but are too small to be resolved in this plot.}
    \label{fig:rffa-cands}
\end{figure*}

Our search for periodic signals from PSR J2021$+$4026 in the TS mode data generated a total of 215 candidates between $262.5 < P < 267.5$ ms. Figure \ref{fig:rffa-cands} shows the period$-$DM distribution of candidates detected from all of the individual 31 minute TS observations as well as the known period range across our dataset (vertical line) and predicted DM (horizontal line).   

Were the pulsar persistently emitting detectable radio emission across all observations we would expect to see a clustering of points located along the vertical dashed lines of Figure \ref{fig:rffa-cands} at a height according to the true DM of the pulsar. Though there are clear clusters of low DM candidates around $P = 0.2635$ and $P=0.2668$ s, they are sufficiently far away from the true period of the pulsar that they can safely be ruled out as compelling candidates. 


It is possible that any radio emission from PSR J2021$+$4026 was only detectable in a subset of observations. This could occur if the intrinsic radio flux density is variable $-$ either on the same long timescale as the $\gamma$-ray flux variations or down to shorter timescales that are comparable to the individual observation lengths. Scintillation, the short-term variations in intensity due to turbulence in the interstellar medium along the line-of-sight to the pulsar, could also render the pulse detectable in only a subset of observations (e.g., \citealt{ric90}).  It is therefore important to individually verify each candidate that is close to the pulsar period at any value of DM. We note however that there are no candidates whose period falls within the expected period range given by the two dashed vertical lines of Figure \ref{fig:rffa-cands}. 

We manually inspected the {\sc riptide} candidate plots for all 251 candidates detected and the ten whose period values are closest to the known period of PSR J2021$+$4026 are listed in Table \ref{Table:candidates}. The S/N of these candidates range from 7.0 (the threshold S/N considered in our analysis) up to 16.4. For each of these candidates, clear pulse-like features appear in the integrated pulse profiles, however these arise due to short, non-periodic bursts of RFI occuring in a small number of subintegrations that were not fully excised by our RFI mitigation techniques (see Section \ref{observations}). In no cases do we find a persistent signal across an observation duration or even across more than a small number of subintegrations.  Furthermore, although we list the dispersion measures of the candidates, there was no clear peak in the S/N at a particular value of the candidate's dispersion measure for any of the candidates listed in Table \ref{Table:candidates}. 

\begin{center}
\begin{table*}

    \caption{The ten periodic emission candidates whose topocentric period values fall closest to the known period of PSR J2021$+$4026 generated by {\sc riptide} using our TS observations. MJD\textsubscript{start} refers to the epoch of the first sample in the filterbank in which the candidate was detected.}
    \begin{tabular}{ c | c | c | r | c | r | r | c |}
    \hline
    Observation & MJD\textsubscript{start} & $P$ (s) & DM (pc cm\textsuperscript{-3}) & Width (turns) & S/N & State  \\ \hline \hline
    
    8 & 58900.164050925923 & 0.265195504 & 446.0 & 0.016064257 & 7.041175365 & D \\
    8 & 58900.164050925923 & 0.265296704 & 328.0 & 0.012048193 & 7.007532597 & D \\
    6 & 56625.855509259258 & 0.265478634 & 252.0 & 0.168674699 & 16.440572739 & B \\
    5 & 56616.165729166663 & 0.265269293 & 82.0 & 0.008032129 & 9.649990082 & B \\
    5 & 56616.165729166663 & 0.265164697 & 10.0 & 0.004016064 & 8.558164597 & B \\
    5 & 56616.318113425928 & 0.265133828 & 24.0 & 0.016064257 & 8.861689568 & B \\
    5 & 56616.209328703706 & 0.265562592 & 22.0 & 0.012048193 & 7.954527855 & B \\
    5 & 56616.209328703706 & 0.265093431 & 8.0 & 0.004016064 & 7.354841232 & B \\
    5 & 56616.209328703706 & 0.265342497 & 8.0 & 0.008032129 & 8.678928375 &  B \\
    5 & 56616.209328703706 & 0.265756531 & 16.0 & 0.008032129 & 8.455500603 & B \\
    5 & 56616.143888888888 & 0.265402513 & 6.0 & 0.008032129 & 9.185052872 & B \\
    \hline
    \end{tabular}
    \label{Table:candidates}
\end{table*}
\end{center}

\begin{figure}
    \centering
    \includegraphics[width=\columnwidth]{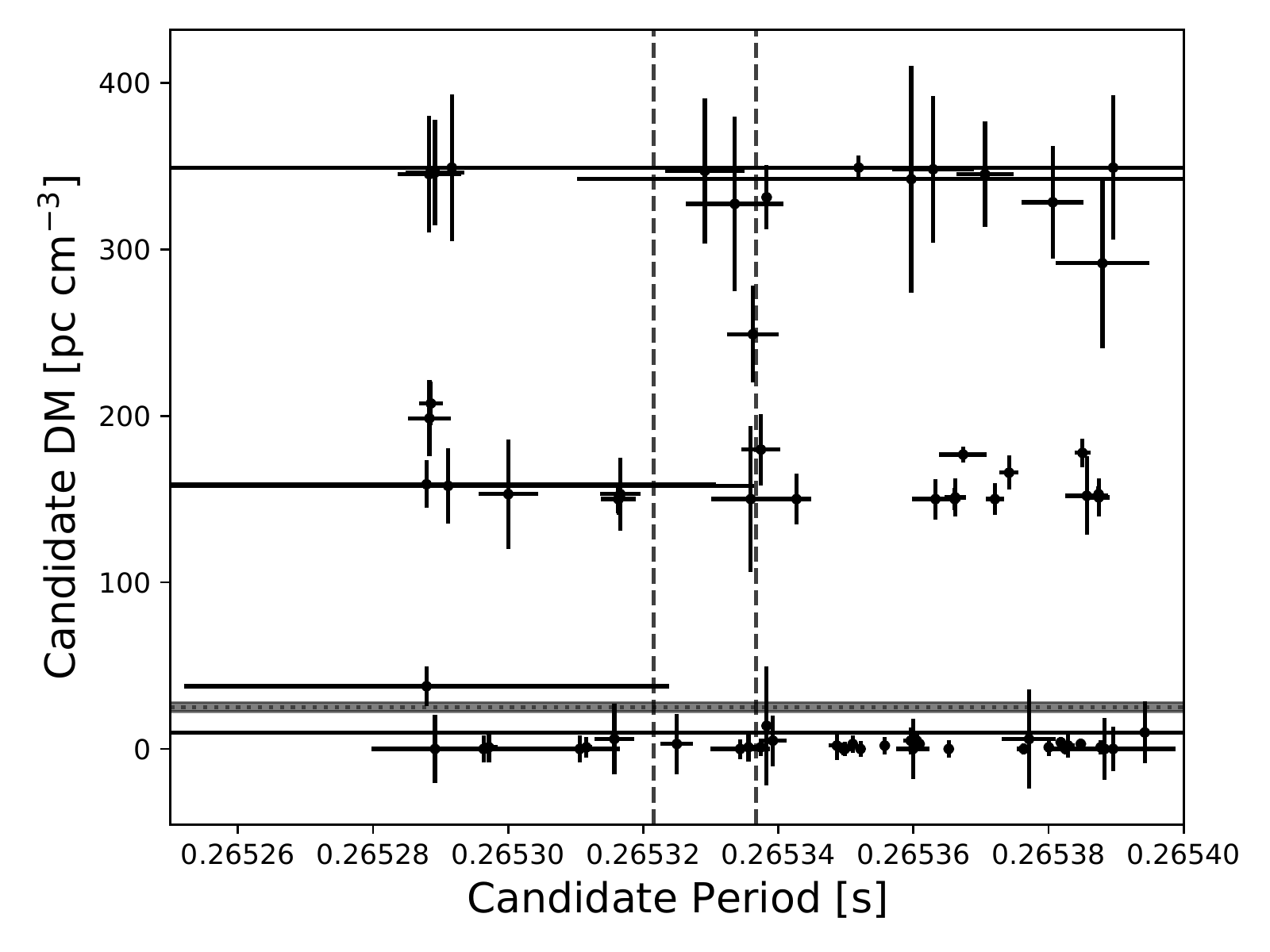}
    \caption{Optimised period and DM values for each of the 2011 PF observations of  PSR J2021$+$4026 calculated using {\sc pdmp}. The errorbars represent the $1\sigma$ uncertainty on the parameters. The vertical and horizontal dashed lines are the same as those shown in Figure \ref{fig:rffa-cands}.}
    \label{fig:pdmp}
\end{figure}

Figure \ref{fig:pdmp} shows the results of our search for an optimal period and DM for each of our PF archives using {\sc pdmp}. For each archive, {\sc pdmp} yields the period and DM combination which maximises the S/N. As each archive was searched twice using distinct but overlapping DM ranges, there are twice the number of data points as PF observations in this plot. We note that the optimised parameters shown predominantly occur in three horizontal "layers" at the values of the upper and lower bounds of the DM ranges searched, suggesting that no truly optimal values of period and DM were found which were common across the PF observations. Our inspection of the diagnostic plots from {\sc pdmp} did not reveal evidence of significant integrated pulses and the "optimised" values produced were the result of local S/N maxima in the period-DM space searched.  We therefore find no evidence for periodic radio pulsations from PSR J2021$+$4026 from any of our 1.5 GHz observations. 

Our search for single pulses using {\sc Heimdall} produced 311440 candidates from our TS data, corresponding to an average of approximately 6 candidates per second of data. Candidate S/Ns ranged from our cut-off value of 7 to $\sim$100.  The {\sc Fetch} classifier was then applied to each of the candidates, yielding a probability (of the candidate pulse being of genuine astrophysical origin) and an a corresponding binary classification score (0 or 1). 206 candidates were assigned a score of 1 by {\sc Fetch} and these were evaluated by manual inspection. Our search revealed no evidence of single pulse emission from PSR J2021$+$4026.

\section{Discussion}
\label{discussion}

We have performed the first targeted search at 1.5 GHz for pulsed radio emission from the mode-switching $\gamma$-ray pulsar PSR J2021$+$4026 using a dataset spanning 9 years. Observations have been conducted whilst the pulsar was in each of its two emission/spin-down modes (HGF/LSD and LGF/HSD). We find no evidence of pulsed radio emission from PSR J2021$+$4026 in either of its emission states.  In other words, the pulsar is \emph{radio quiet} at this frequency, at least along the line-of-site towards the Earth. We must distinguish here between radio quiescence and radio silence as we cannot rule out that the pulsar is emitting low-level radio pulsations with a flux density below our sensitivity limits.  The minimum sustained pulsed flux density from PSR J2021$+$4026 that would have been detectable in these observations can be estimated using the modified radiometer equation, 

\begin{equation}
    \label{radiometer}
    S_{\mathrm{min}} = \beta \frac{\mathrm{(S/N)_{min}} T_{\mathrm{sys}}}{G \sqrt{n_\mathrm{pol} \tau_{\mathrm{obs}} \Delta f}} \sqrt{\frac{W}{P - W}},
\end{equation}

\noindent where $T_{\mathrm{sys}} = 44$ K is the system temperature (comprising the sum of the receiver temperature $T_{\mathrm{rec}} = 25$ K and the sky temperature\footnote{Estimated using {\sc pygdsm}. See https://github.com/telegraphic/pygdsm} $T_{\mathrm{sky}} = 19$ K \citep{ztd+17}), $G = 0.6$ K Jy\textsuperscript{-1} is the gain of the instrument, $n_\mathrm{pol} = 2$ is the number of polarisation, $\tau_{\mathrm{obs}}$ is the observation integration time (typically $\sim$1800 s), $\Delta f$ is the maximum bandwidth (384 MHz), $P$ is the period of the pulsar and $W$ is the pulse width (we assume this to be 5\% of one pulse period) and $\beta = 1$ is a factor which corrects for signal digitisation imperfections. In order to estimate a minimum detection S/N threshold, we use {\sc filtools} to inject artificial pulsar-like signals, with parameters corresponding to those of PSR J2021+4026, into our TS filterbank files, and attempt to re-detect those signals using our search pipeline.  We find we are able to re-detect simulated pulsars if the S/N exceeds $\approx$ 12. We therefore adopt this as the minimum detection threshold in Equation \ref{radiometer}.  Using these values, we estimate an upper limit of the L-BAND flux density of PSR J2021$+$4026 $S_\mathrm{min} \approx 0.2$ mJy.

Targeted searches at other frequencies have also been conducted. \cite{bwa+04} searched for radio emission, using the Green Bank telescope (GBT), from the $\gamma$-ray source 3EG J2020+4017 prior to its identification as PSR J2021$+$4026. No pulsed radio emission was detected and they calculated upper limits of the 820 MHz flux density of 40 $\mu$Jy. However, this was computed before a precise position could be determined for the pulsar, therefore this value may be overestimated. \cite{rkp+11}, also using GBT, similarly did not detect radio emisson from PSR J2021$+$4026 at 2 GHz, establishing a flux density upper limit of 11 $\mu$Jy. We note that no radio flux limits have been determined for this pulsar at lower frequencies.

New generation radio telescopes will be particularly useful for the study of neutron stars in apparent radio quiescence. With dramatically improved sensitivities and ultra-wide bandwidths, facilities such as the Square Kilometre Array (e.g., \citealt{tkb+15}, \citealt{kjk+15}) and the Five hundred metre Aperture Spherical Telescope (e.g., \citealt{lw+18}) will provide the deepest radio searches to-date for sources such as PSR J2021$+$4026, establishing them as faintly emitting radio pulsars or providing improved constraints on their radio flux densities. Such searches may reveal PSR J2021$+$4026 to be faintly radio emitting in one or both of its spin-down states. If the former, PSR J2021$+$4026 would join the class of intermittent pulsars such as PSR B1931+24 (\citealt{klo+06}) whose radio flux is currently only detectable when the pulsar assumes one of its two emission states.  

Alternatively, with deeper searches, the pulsar may be shown to be emitting detectable radio emission in both spin-down/$\gamma$-ray flux states, exhibiting either spin-correlated radio mode-switching behaviour (e.g., \citealt{lhk+10}, \citealt{ssw+22}) in addition to $\gamma$-ray mode-switching, or stable radio emission with no detectable changes to its pulse shape or flux.  \cite{tim10} suggested a model in which the structure of the pulsar magnetosphere changes in such a way that region encompassing closed magnetic field lines can assume different sizes according to a particular magnetospheric state. The line-of-sight therefore, cuts a different path across the beam in each state and this would manifest as an observed mode-switch. Alternatively, in edge cases, the beam may only crosses the line-of-sight in its wider state, leading to observed intermittency. 

The relative widths of $\gamma$-ray profiles suggest an high-altitude emission region in the outer magnetosphere, near the light cylinder (see \citealt{har18} and references therein for a review of pulsar emission models), whereas radio emission is expected to be generated near the polar cap, resulting in a narrower beam.  It is therefore expected that some $\gamma$-ray loud pulsars are intrinsically radio-loud but not observable as such, as the radio beam does not intersect our line-of-sight. To-date there is no comprehensive theory that explains the breadth of phenomena observed in pulsar emission across the electromagnetic spectrum. The radio emission is coherent and emitted from a narrow region over the polar cap but the exact conditions that give rise to coherence are not well understood.  The discovery of a radio-emitting $\gamma$-ray mode-switching pulsar would yield valuable insights into the inter-dependency of these different emission regions and the physics that underpins pulsar emission at multiple wavelengths. 

Although PSR J2021$+$4026 is the only $\gamma$-ray pulsar observed to exhibit mode-switching, a small number of X-ray emitting radio pulsars are known to exhibit this phenomenon. The $P = 1.1$ s, $\tau_\mathrm{char} = 5$ Myr pulsar PSR B0943$+$10 undergoes simultaneous X-ray and radio mode-switching, characterised by thermal X-ray pulsations exhibiting a greater flux when the radio flux is lower (\citealt{mkt+16}, \citealt{rmt+19}). PSR B0826$+$26 ($P = 530$ ms) has a similar $\tau_\mathrm{char}$ to PSR B0943$+$10 and also undergoes mode-switches on a similar timescale but in this case, the high-flux radio state is coincident with the high-flux X-ray state \citep{hkb+18}. The younger $P = 0.7$ s radio mode-switching pulsar PSR B1822$-$09 exhibits no detectable changes to its pulsed X-ray emission \citep{hkh+17} but clear rapid changes to the radio pulse profile. Due to the small magnitudes of pulsar spin-down rates it is difficult to detect coincident spin-down transitions on these short timescales (e.g., \citealt{ssw18}) however spin-down variations have been linked to longer term profile modulations in PSR B1822$-$09 \citep{lhk+10} and PSR B0943$+$10. As a complicated picture is emerging across this small handful of high-energy mode-switching pulsars, it is clear that a larger sample of variable X-ray and $\gamma$-ray sources is essential for understanding the complex relationships between the different pulsar emission behaviours at all wavelengths, including deep radio searches for apparently radio-quiet $\gamma$-ray pulsars.   

Stochastic variations in the density of the ISM along the line-of-sight gives rise to intensity variations in pulsar radio emission (scintillation) on timescales of minutes to hours.  Assuming the pulsar is beaming radio emission towards Earth, we can estimate the impact of interstellar scintillation on our likelihood of detecting radio pulsations from PSR J2021$+$4026 using the same Cordes-Lazio NE2001 model used to estimate the pulsar's DM\footnote{see https://www.nrl.navy.mil/rsd/RORF/ne2001/} \citep{cl2003}. For a distance of 2.15 kpc at the position of PSR J2021$+$4026, we find the estimates of scintillation bandwidth $\Delta f_{\mathrm{scint}} \sim$36 MHz and timescale $\tau_{\mathrm{scint}} \sim$1400 \nolinebreak s ($\sim$1\% of the duration of all of our observations combined). Although $\tau_{\mathrm{scint}}$ covers $\sim$70\% of a typical 31 minute observation, the predicted $\Delta f_{\mathrm{scint}}$ cover only $\sim$10\% of our observing band. It is therefore unlikely that scintillation is responsible for the radio non-detection of PSR J2021$+$4026. 

Upon the discovery of the first transition in PSR J2021$+$4026, \cite{abb+13} noted a $\sim$6\% change to the spin-down rate of the star and commented on its \emph{glitch-like} nature. Glitches are sudden, discontinuous increases to the rotation rate of a pulsar. Over time (hours to years)  a pulsar may undergo a period of recovery from a glitch in which the rotation rate quasi-exponentially relaxes towards the value it would have reached by that time, without the occurrence of the glitch. In many cases, particularly in younger pulsars (which are statistically more likely to undergo glitches), changes to the $\dot{\nu}$ are also observed (see e.g., \citealt{fer+17} and \citealt{bsa+22} for an overview of glitch statistics). Glitches are typically understood to be a manifestation of angular momentum redistribution between an interior ocean of superfluid neutrons and the stellar crust, possibly mediated by tectonic activity in the crusts of young neutron stars (see e.g., \citealt{hm15}, \citealt{zgy+22}). 

Spin-down transitions are commonly observed in young to middle-aged pulsars and may be at least partially responsible for timing noise (see e.g., \citealt{hlk+04}, \citealt{bkj+15}, \citealt{pjs+20}, \citealt{ssw+22}).  A number of radio pulsars are known to transition quasi-periodically between (typically two) otherwise stable spin-down rates, in some cases accompanied by radio emission changes (e.g., \citealt{lhk+10}, \citealt{ssw+22}). The timescales of spin-down transitions can be difficult to examine given the quadratic signature they introduce into pulsar timing residuals, especially where pulse time-of-arrival (TOA) uncertainty is large and/or observing cadence is low. This can lead to difficulties in distinguishing between smaller glitches and spin-down transitions \citep{ssw18}. Nevertheless, these events are typically regarded as distinct from glitches and although the exact mechanism that underpins spin-down and/or emission changes is not fully established, they are understood to arise from changes to the distribution of magnetospheric charges which in turn affect the pulsar's braking torque (e.g., \citealt{klo+06}).

Following \cite{abb+13}, a number of authors (e.g., \citealt{ntc16}, \citealt{znl+17}, \citealt{wth+18}) have speculated that the 2011 increase in $|\dot{\nu}|$ may be symptomatic of a glitch event. For example, \cite{ntc16} suggested that glitch related crust cracking could result in a shift in the magnetic inclination angle, affecting the $\gamma$-ray profile and $\dot{\nu}$, resulting in the variability observed. The inclination angle of the Crab pulsar (PSR B0531$+$21) has been observed to be increasing over time \citep{lgw+13}, however this is a secular variation and is uncorrelated with the prolific glitch activity of the Crab pulsar.  \cite{znl+17} showed that local changes to the polar cap geometry due to a glitch could trigger a change to the magnetospheric current density, resulting in an observed mode-switch and coincident change to $\dot{\nu}$, and that the subsequent return of $\dot{\nu}$ to its pre-2011 value may be a manifestation of post-glitch relaxation.  We note however that, of the $\sim$600 glitches reported in the Jodrell Bank Glitch Catalogue\footnote{see https://www.jb.man.ac.uk/$\sim$pulsar/glitches/gTable.html} only a very small number have been associated with radiative activity.  For example, glitches in the highly magnetised radio pulsar PSR J1119$-$6127 have been accompanied by magnetar-like X-ray activity and periods of erratic post-glitch profile evolution (e.g., \citealt{wje11},  \citealt{awe+15}, \citealt{wld+20}).  The Vela pulsar (PSR B0833$-$45) exhibited some transient changes to its radio profile shape very close in time to its December 2016 glitch \citep{pdh+18}. There is also some evidence that the correlated radio emission and $\dot{\nu}$ variability of PSR B0740$-$28 are influenced by glitch events \citep{ksj13}. Permanent, or semi-permanent changes to pulsed emission due to glitch activity has not been observed. That said, canonical glitches have been observed in the related radio-quiet $\gamma$-ray pulsar, Geminga \citep{jhgm02} which is somewhat older that PSR J2021$+$4026. As glitch activity is anti-correlated with age, we may expect to observe future glitches in PSR J2021$+$4026. 

As the population of known pulsars grows and long-term timing programs accrue ever-longer time baselines, the number of pulsars observed to undergo $\dot{\nu}$ transitions that are coincident with emission changes is increasing (e.g., \citealt{klo+06}, \citealt{lhk+10}, \citealt{bkb+14}, \citealt{psw14}, \citealt{bkj+15}, \citealt{psw+16}, \citealt{ssw+22}). The $\dot{\nu}$-correlated $\gamma$-ray variability of PSR J2021$+$4026 bears a remarkable resemblance to the similar bimodal transitioning seen in many radio pulsars and we suggest that a similar magnetospheric switching mechanism underpins its variability. As the first (and so far only) $\gamma$-ray pulsar to show such correlated state-switching behaviour it offers a valuable opportunity to study the dynamic conditions in pulsar magnetospheres. 


\begin{figure}
    \centering
    \includegraphics[width=1.0\columnwidth]{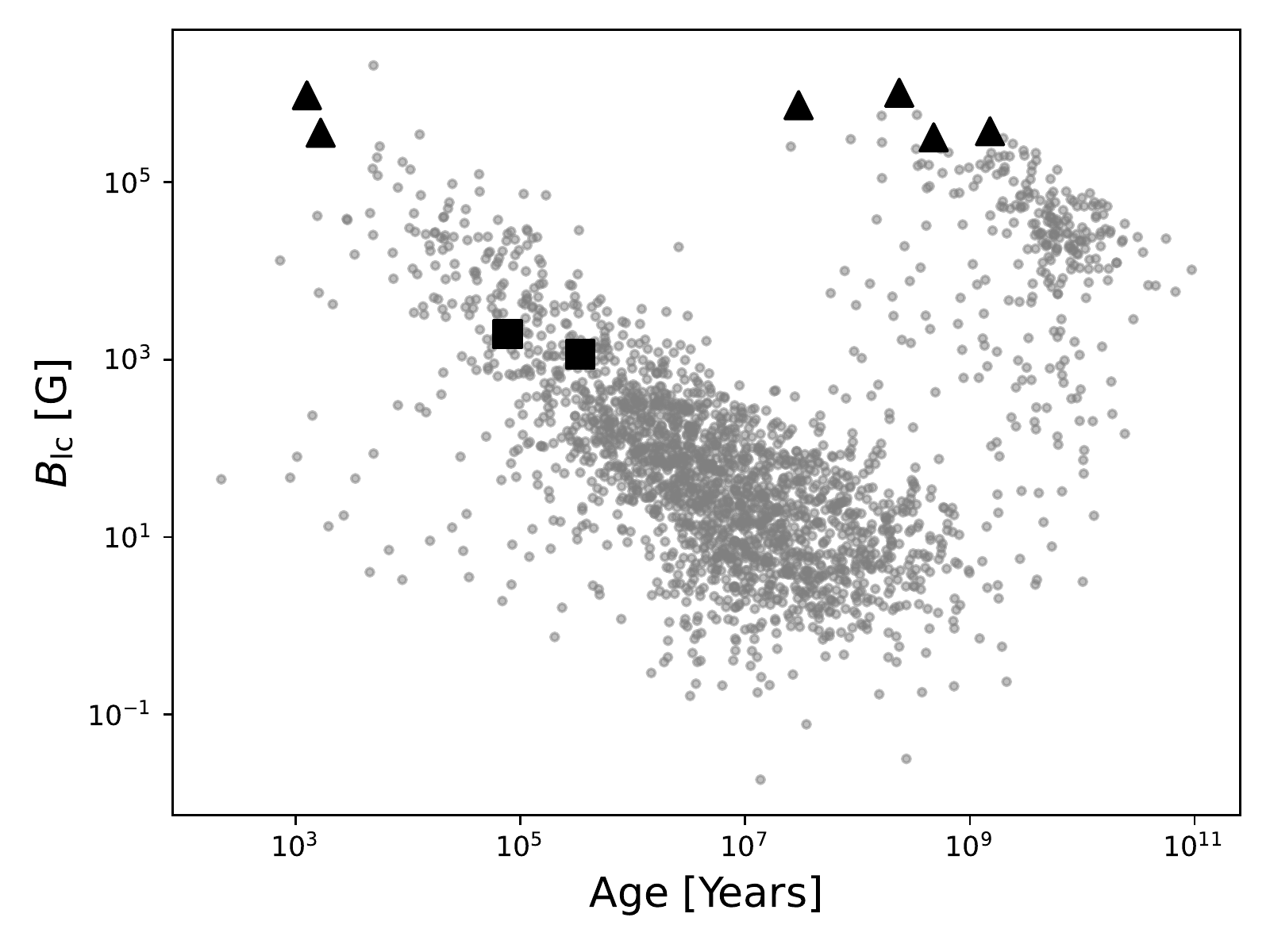}
    \caption{Magnetic flux density at the light cylinder versus the characteristic age of all of the pulsars (small gray circles) in the ATNF pulsar catalogue where entries exist for these quantities \citep{mhth05}. The black triangles represent pulsars from which giant radio pulses have been observed. The two black squares are PSR J2021$+$4026 (leftmost) and Geminga.}
    \label{fig:blc}
\end{figure}

We did not detect any transient radio emission from PSR J2021$+$4026 throughout our observations, in the form of either RRAT-like behaviour or GP emission.  The lack of GPs is perhaps not a surprising result given the relatively modest value of the magnetic flux density at the light cylinder $B_\mathrm{lc}$ of this pulsar ($\sim$ 2000 G). Figure \ref{fig:blc} shows the values of $B_\mathrm{lc}$ vs the characteristic age $\tau_\mathrm{char}$ for the pulsar population. The distribution is broadly divided into two islands corresponding to the bulk of the population, typically termed `normal' pulsars and the millisecond pulsars (upper right corner). The gray triangles represent pulsars known to emit GPs and these sources consistently have values of $B_\mathrm{lc}$ that are 2 - 3 orders of magnitude greater than PSR J2021$+$4026, represented by the leftmost black square.  The other black square represents Geminga, a radio quiet $\gamma$-ray pulsar with similar properties to PSR J2021$+$4026, also from which no GPs have been observed. 

Typically, a targeted search for GPs would require a sampling interval that is shorter than the expected pulse width. As our sampling interval was 256 $\mu$s and GPs typically occupy $\sim$10 $-$ 100 ns, any giants pulses observed from PSR J2021$+$4026 would have likely been unresolved. We can estimate the minimum flux density $S_\mathrm{min}$ for a GP to be detectable in these observations using the radiometer equation. For a GP with a width $\tau_\mathrm{pulse}$ equal to the sampling interval $\tau_\mathrm{samp}$,

\begin{equation}
    \label{radiometer_gp}
    S_{\mathrm{min}} = \beta \frac{\mathrm{(S/N)_{min}} T_{\mathrm{sys}}}{G \sqrt{n_\mathrm{pol} \tau_{\mathrm{samp}} \Delta f}},
\end{equation}


\noindent which yields a $12\sigma$ detection threshold of $\approx$ 2.0 Jy. All other parameters are the same as those used for Equation \ref{radiometer}. Assuming a typical value for $\tau_\mathrm{pulse} = 100$ ns, which is 2560 times shorter than $\tau_\mathrm{samp}$ we find the minimum flux density for a GP to be detected with $12\sigma$ confidence is $2.0 \times \sqrt{2560} \approx 100$ Jy, therefore we are insensitive to any GPs emitted by PSR J2021$+$4026 with a flux density that is less than this value. Applying this method to determine our sensitity to single pulse due to RRAT-like behaviour, assuming any sporadic RRAT-like pulses from J2021$+$4026 have a $\tau_\mathrm{pulse}$ of 5\% of one pulse period, (approximately 50 times longer than $\tau_\mathrm{samp}$), the minimum single pulse flux density for the detection of an RRAT burst is $2.0\times \sqrt{(1/50)} \approx 0.3$ Jy. 

\section{Conclusions}

Using observations spanning over 9 years, we have conducted the first 1.5 GHz search for radio emission from the radio-quiet variable $\gamma$-ray pulsar PSR J2021$+$4026. Our dataset includes observations of the pulsar undertaken during both of the quasi-stable spin-down/emission states between which the pulsar transitions on a timescale of 3 $-$ 4 years. We have used new techniques to search for periodic radio emission at and near the known rotation period of PSR J2021$+$4026. We have also searched for single pulse radio emission, characteristic of giant radio pulses and sporadic emission from rotating radio transients. In neither case do we find any compelling candidates for radio emission from PSR J2021$+$4026 in either of its emission/spin-down states.  We have estimated upper limits of the 1.5 GHz flux density of PSR J2021$+$4026 for periodic, giant pulsed and RRAT-like pulsed emission of 0.2 mJy, 100 Jy and 0.3 Jy respectively. 

The absence of detectable radio emission implies either 
\begin{itemize}
    \item PSR J2021$+$4026 is radio-quiet but not radio-silent and its radio flux density is below our sensitivity threshold in both of its emission/spin-down states,
    \item PSR J2021$+$4026 has an active radio beam that does not cross the line-of-sight to the Earth in either of its emission/spin-down states, 
    \item PSR J2021$+$4026 is not a radio pulsar. 
\end{itemize}

The transitioning behaviour of this pulsar between distinct quasi-stable $\gamma$-ray emission/spin-down states at intervals of several years bears a striking resemblance to the similar phenomenon seen in many radio pulsars. The variability in this pulsar as seen in $\gamma$-rays can be explained by changes to the magnetospheric conditions which results in the observed profile variations whilst also affecting the braking torque on the pulsar. This scenario is our preferred interpretation of the variability of PSR J2021$+$4026. If this is the case, we would expect the pulsar to undergo further transitions in the future and we emphasise that observational resources operating at radio wavelengths should be directed at PSR J2021$+$4026 when these transitions occur, as informed by ongoing  $\gamma$-ray monitoring. As new ever-more sensitive radio facilities come online, the pulsar may be detected as a radio pulsar in one or both of its spin-down/emission states, affording a unique opportunity to study a transitioning pulsar magnetosphere across the electromagnetic spectrum.

\section*{Data Availability}

The data underlying the work in this manuscript are available upon reasonable request.

\section*{Acknowledgements}

The authors thank Dr. Lina Levin, Dr. Vincent Morello and Dr. Kaustubh \mbox{Rajwade} for useful discussions, and the referee for providing useful comments which improved the quality and clarity of this manuscript.   Pulsar research at Jodrell Bank is supported by a consolidated grant from the UK Science and Technology Facilities Council (STFC). 




\bibliographystyle{mnras}
\bibliography{journals,modrefs,psrrefs} 






\bsp	
\label{lastpage}
\end{document}